\documentclass[journal abbreviation]{copernicus2}

\begin{document}


\title{Magnetic field generation in a jet-sheath plasma via the kinetic Kelvin-Helmholtz instability}

\author[1]{K.-I. Nishikawa}
\author[2]{P. Hardee}
\author[3]{B. Zhang}
\author[4]{I. Du\c{t}an}
\author[5]{M. Medvedev}
\author[6]{E. J. Choi}
\author[6]{K. W. Min}
\author[7]{J. Niemiec}
\author[8]{Y. Mizuno}
\author[9]{A. Nordlund}
\author[9]{J. T. Frederiksen}
\author[10]{H. Sol}
\author[11]{M. Pohl}
\author[12]{D. H. Hartmann}

\affil[1]{Center for Space Plasma and Aeronomic Research,
University of Alabama in Huntsville, 320 Sparkman Drive, ZP12,
Huntsville, AL 35805, USA}
\affil[2]{Department of Physics and Astronomy,
  The University of Alabama, Tuscaloosa, AL 35487, USA}
\affil[3]{Department of Physics, University of Nevada, Las
Vegas, NV 89154, USA}
\affil[4]{Institute of Space Science, Atomistilor 409, Bucharest-Magurele RO-077125, Romania}
\affil[5]{Department of Physics and Astronomy, University of Kansas, KS
66045, USA}
\affil[6]{Korea Advanced Institute of Science and Technology, 
Daejeon 305-701, South Korea}
\affil[7]{Institute of Nuclear Physics PAN, ul. Radzikowskiego 152, 31-342 Krak\'{o}w, Poland}
\affil[8]{Institute of Astronomy
National Tsing-Hua University,
Hsinchu, Taiwan 30013, R.O.C}
\affil[9]{Niels Bohr Institute, University of Copenhagen, 
Juliane Maries Vej 30, 2100 Copenhagen \O, Denmark}
\affil[10]{LUTH, Observatore de Paris-Meudon, 5 place Jules Jansen, 92195 Meudon Cedex, France}
\affil[11]{Institue of Physics and Astronomy, University of Potsdam, Karl-Liebknecht-Strasse 24/25
14476 Potsdam-Golm, Germany}
\affil[12]{Department of Physics and Astronomy, Clemson University, Clemson, SC 29634, USA}


\runningtitle{Magnetic field generation in jet-sheath plasma}

\runningauthor{Nishikawa et al.}

\correspondence{K.-I. Nishikawa\\ (ken-ichi.nishikawa@nasa.gov)}

\received{}
\pubdiscuss{} 
\revised{}
\accepted{}
\published{}


\firstpage{1}

\maketitle  

\begin{abstract}
WWe have investigated generation of magnetic fields associated with velocity shear between an unmagnetized relativistic jet and  an unmagnetized sheath plasma.  We have examined the strong magnetic fields generated by kinetic shear (Kelvin-Helmholtz) instabilities. Compared to the previous studies using counter-streaming performed by Alves et al. (2012), the structure of KKHI  of our jet-sheath configuration is slightly different even for the global evolution of the strong transverse magnetic field. In our simulations the major components of growing  modes are the electric field $E_{\rm z}$ and the magnetic field $B_{\rm y}$. After the $B_{\rm y}$ component is excited, an induced electric field $E_{\rm x}$ becomes significant. However, other field components remain small. We find that the structure and growth rate of  KKHI  with mass ratios $m_{\rm i}/m_{\rm e} = 1836$ and $m_{\rm i}/m_{\rm e} = 20$ are similar. In our simulations  saturation in the nonlinear stage is not as clear as in  counter-streaming cases. The growth rate for a  mildly-relativistic jet case ($\gamma_{\rm j} = 1.5$) is larger than for a  relativistic jet case ($\gamma_{\rm j} = 15$).  
\end{abstract}


\introduction  
Recent kinetic simulations have focused on magnetic field generation via electromagnetic plasma instabilities in unmagnetized flows without velocity shear. Three-dimensional (3D) particle-in-cell (PIC) simulations of Weibel 
turbulence (Nishikawa et al. 2005, 2008, 2009a) have demonstrated subequipartition magnetic field generation.

Velocity shears also must be considered when studying particle acceleration scenarios, since these trigger the kinetic 
Kelvin-Helmholtz instability (KKHI). In particular the KKHI has been shown to lead to particle acceleration and 
magnetic field amplification in relativistic shear flows (Alves et al. 2012;  Liang et al. 2013; Nishikawa et al. 2013). 
Furthermore, a shear flow upstream of a shock can lead to density inhomogeneities via the MHD Kelvin-Helmholtz 
instability (KHI) which may provide important scattering sites for particle acceleration.

\begin{figure*}[ht!]
\begin{center}
\resizebox{0.70\hsize}{!}{
\includegraphics[width=12cm]{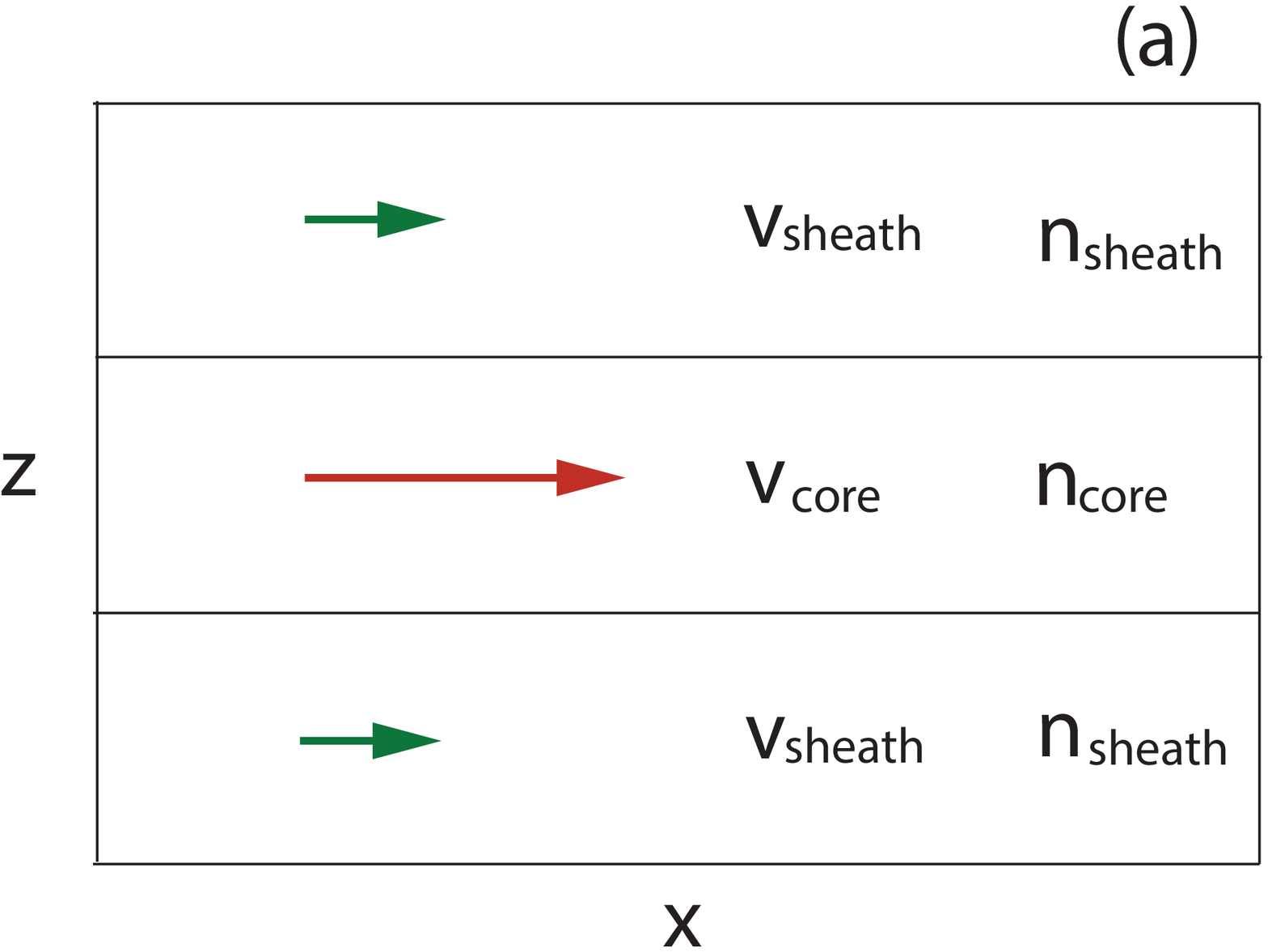}
\quad
\includegraphics[width=10cm]{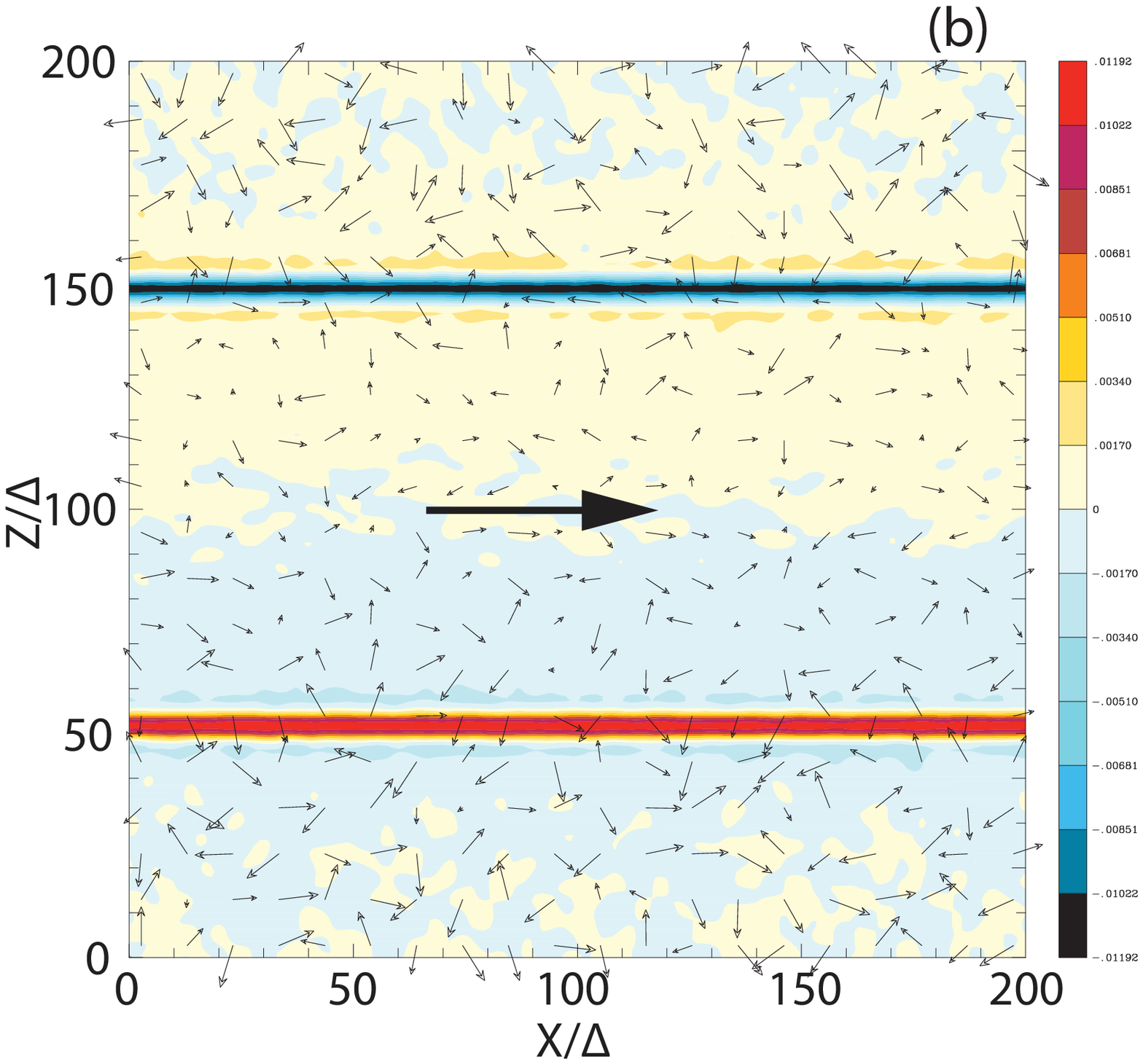}} 
\end{center}
\vspace*{-0.5cm}
\caption{\baselineskip 12.0pt Figure \ref{fig1}a  shows our simulation model where the sheath plasma can be stationary or moving in the same direction as the jet core. In this simulation the sheath velocity is zero.   Figure \ref{fig1}b shows the magnitude of $B_{\rm y}$  is plotted in the $x - z$ plane (jet flow in the $+x$-direction indicated by the large arrow) at the center of the simulation box, $y = 100\Delta$ at simulation time $t = 70\,\omega_{\rm pe}^{-1}$ for the case of  $\gamma_{\rm j}=15$ and $m_{\rm i}/m_{\rm e} = 1836$. This panel covers one fifth of the whole simulation system. The arrows show the magnetic field in the plane.
\label{fig1}}
\end{figure*}

In our simulations a relativistic jet  plasma is surrounded by a sheath plasma (Nishikawa et al. 2013).  This setup is 
similar to the setup of our RMHD simulations \citep{mizuno07}. In one of our simulations, the jet core has 
$v_{\rm core} = 0.9978c ~(\gamma_{\rm j}=15)$ pointing in the positive $x$ direction in the middle of the simulation box 
as in Alves et al. (2012). Unlike Alves et al. (2012)\ the upper and lower quarter of the simulation box contain a stationary, 
$v_{\rm sheath}= 0$, sheath plasma as shown in Fig. \ref{fig1}a. Our setup allows for motion of the sheath plasma
 in the positive $x$ direction. 

\begin{figure*}[ht!]
\begin{center}
\resizebox{.70\hsize}{!}{\includegraphics[width=36cm]{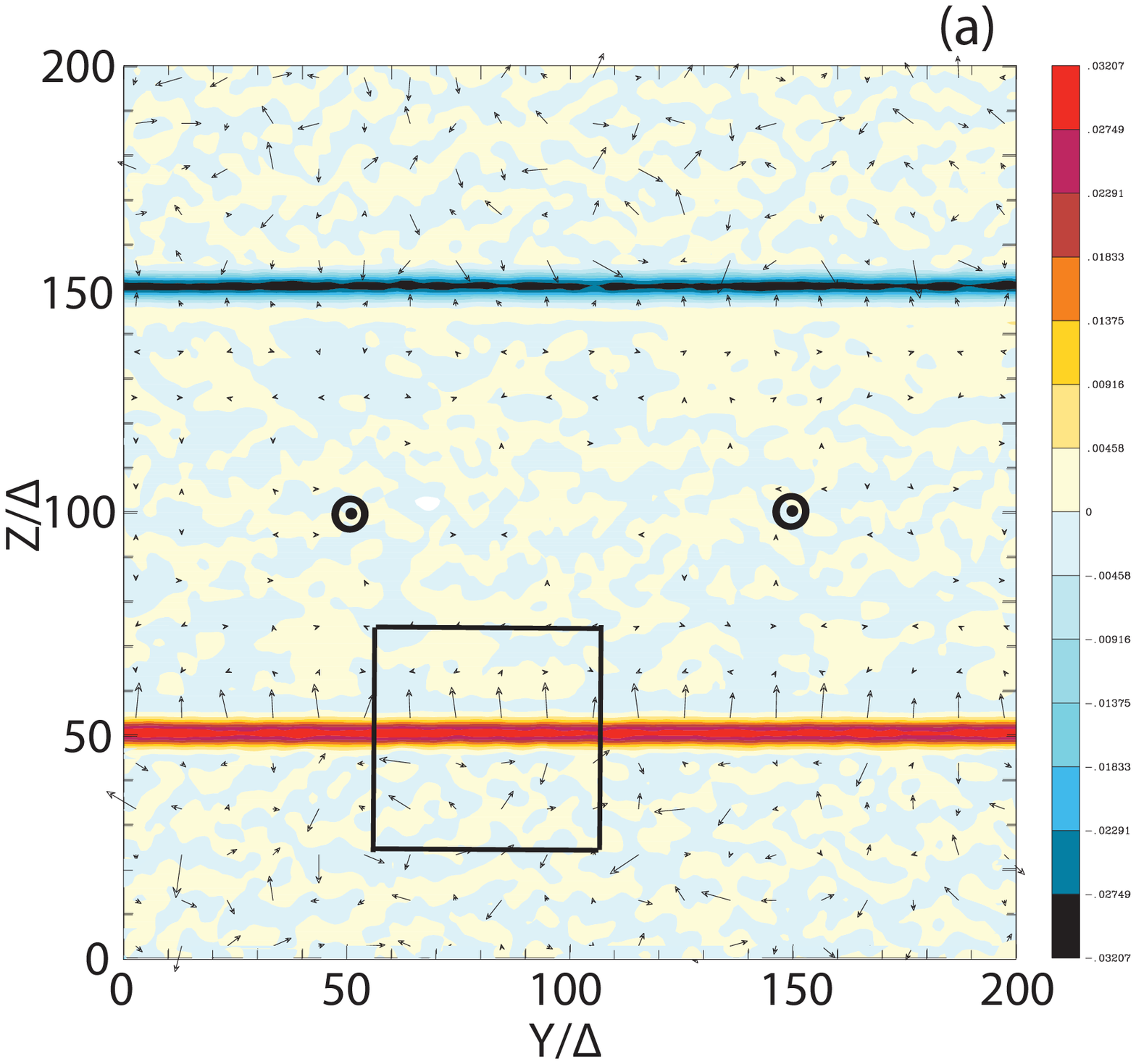}
\quad
\includegraphics[width=20cm]{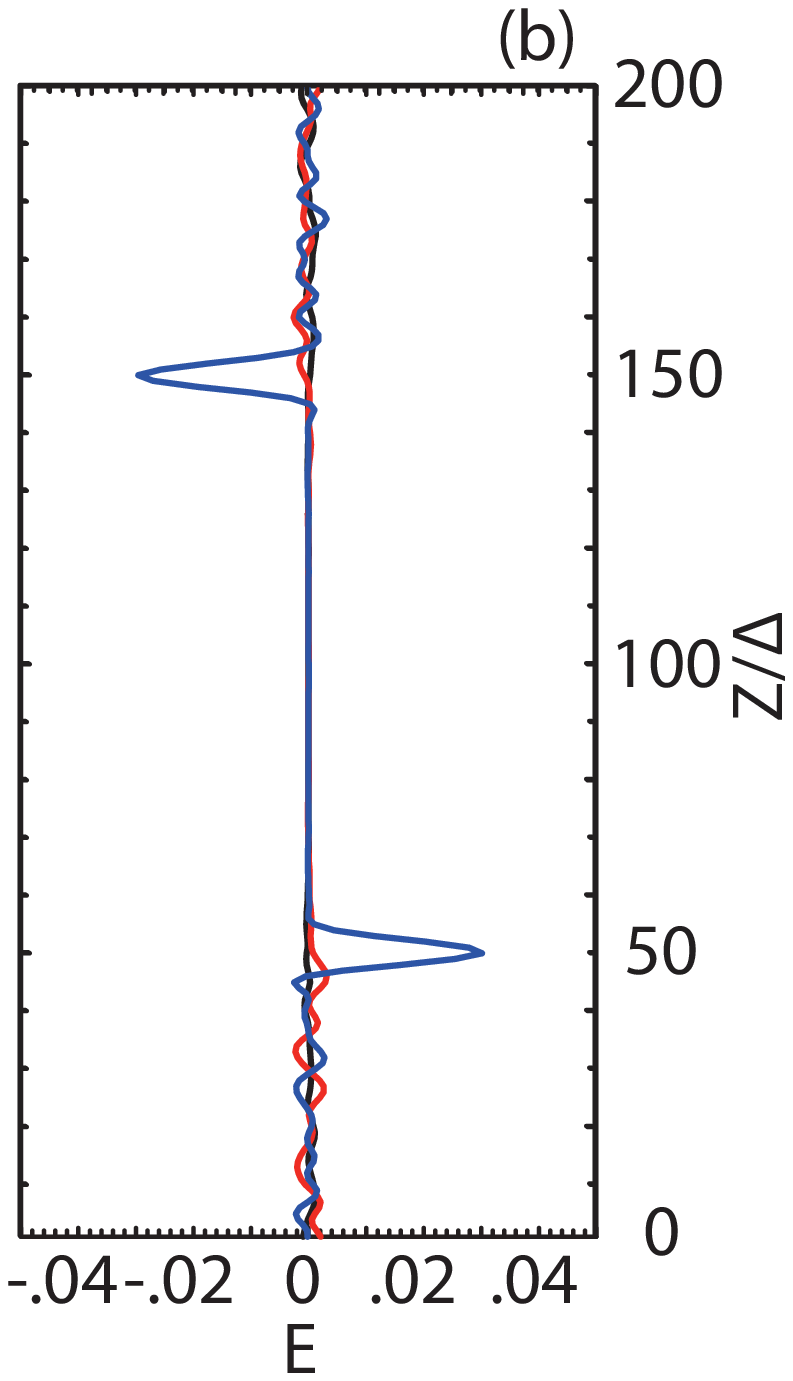} 
\,
\includegraphics[width=21cm]{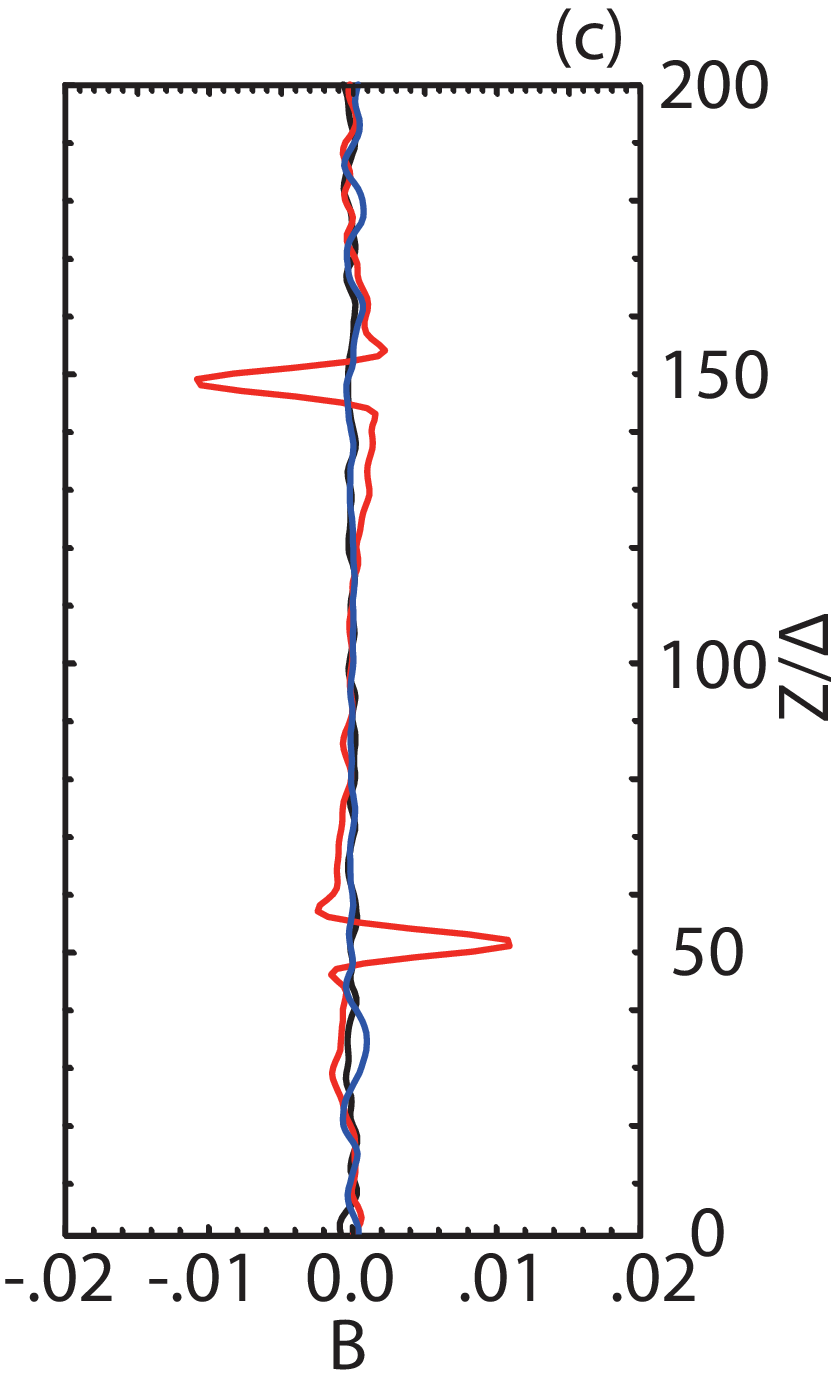}}
\end{center}
\vspace*{-0.5cm}
\caption{\baselineskip 12.0pt Electric and magnetic field generated by a relativistic electron ion jet core with 
$\gamma_{\rm j} = 15$ and stationary sheath plasma for the case with the mass ratio $m_{\rm i}/m{\rm e} = 1836$. 
The magnitude of  $E_{\rm z}$ is plotted in the $y - z$ plane (jet flow is out of the page) at the center of the 
simulation box,  at $x = 500\Delta$ at $t = 30 \omega_{\rm pe}^{-1}$ (panel (a)).  Panel (b) shows $E_{\rm z}$ (blue), 
$E_{\rm x}$ (black), , and $E_{\rm y}$ (red) at $x = 500\Delta$ and $y = 100\Delta$ at the same time. 
Since the magnetic field grows more slowly (see Fig. \ref{fig4}), panel (c) shows   $B_{\rm y}$ (red), $B_{\rm x}$
 (black), , and $B_{\rm z}$ (blue) at  $t = 70 \omega_{\rm pe}^{-1}$.  
\label{fig2}}
\end{figure*}

Overall, this structure is similar in spirit, although not in scale, to that proposed for active galactic nuclei (AGN) 
relativistic jet cores surrounded by a slower moving sheath, and is also relevant to gamma-ray burst (GRB) jets. 
In particular, we note that this structure is also relevant to the ``jet-in-a-jet" or ``needles in a jet" scenarios
\citep[and papers therein]{giannios09}, which have been invoked to provide smaller scale high speed structures 
within a much larger more slowly moving AGN jet. Similar smaller scale structures within GRB jets are also conceivable.

This more realistic setup 
will allows us to compute synthetic spectra in the observer frame.
As mentioned by Alves et al. (2012), in a non-counter-streaming or unequal density counter-streaming setup the growing KKHI will propagate with the flow. For GRB jets, the relativistic jet core is thought to have a  much higher density
relative to the external medium. On the other hand, for an AGN jet the relativistic core is thought to be  less dense than the surrounding sheath.

Recently the KKHI was investigated using  a  counter-streaming velocity shear setup 
for  subrelativistic $\gamma_{\rm 0} = 1.02$ and  relativistic  $\gamma_{\rm 0} = 3$ cases,
 (Alves et al.\  2012).   The shear flow initial condition was set by a velocity field with $v_{0}$ pointing
in the positive $x_{1}$ direction, in the upper and lower quarters of the simulation box, and a symmetric velocity field 
with $-v_{0}$  pointing in the negative $x_{1}$ direction, in the middle-half of the box. Initially, the systems were 
charge and current neutral. In the subrelativistic case, the simulation box dimensions were $20\times20\times20(c/\omega_{\rm p})^{3}$, where  $\omega_{\rm p} = (4\pi ne^{2}/m_{\rm e})^{1/2}$ is the plasma frequency, 
and they used 20 cells per electron  skin depth ($c/\omega_{\rm p}$). The simulation box dimensions for the relativistic 
scenario were $250 \times 80 \times 80(c/\omega_{\rm p})^{3}$, with a resolution of 4 cells per ($c/\omega_{\rm p}$).
Periodic boundary conditions were imposed in every direction.

\begin{figure*}[ht!]

\begin{center}
\resizebox{.85\hsize}{!}{\includegraphics[width=18cm]{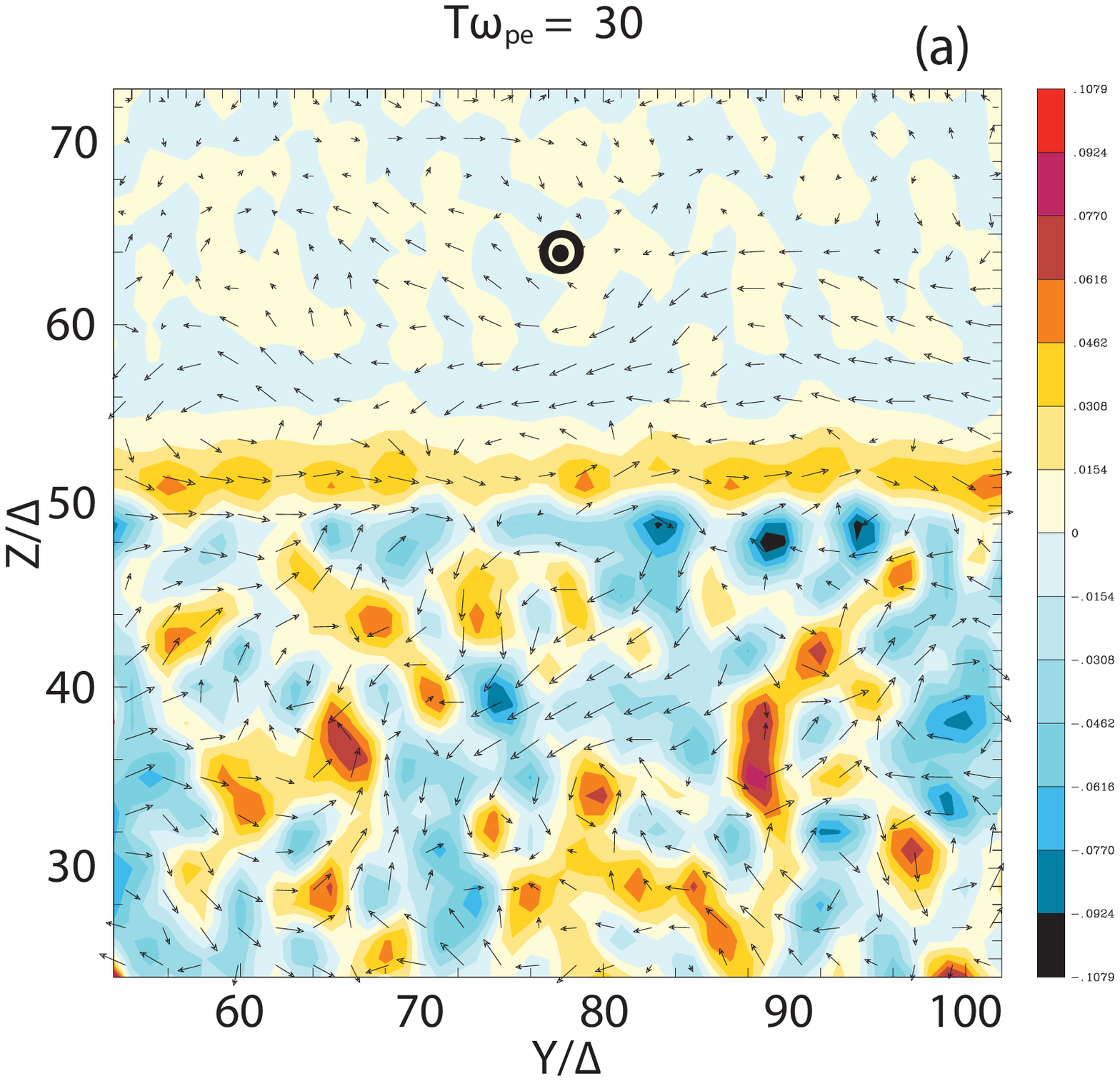}
\,
\includegraphics[width=18cm]{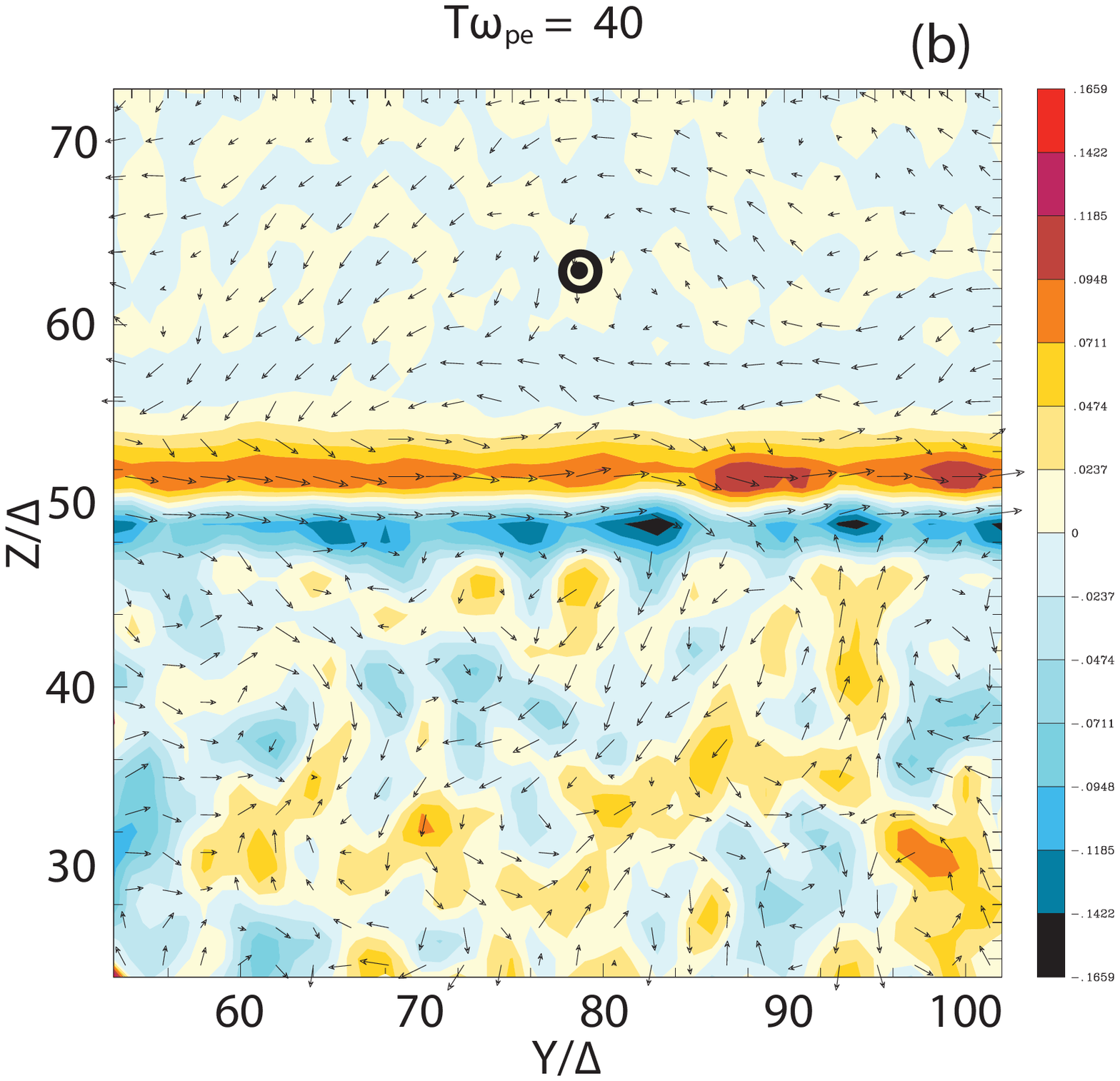} 
\,
\includegraphics[width=18cm]{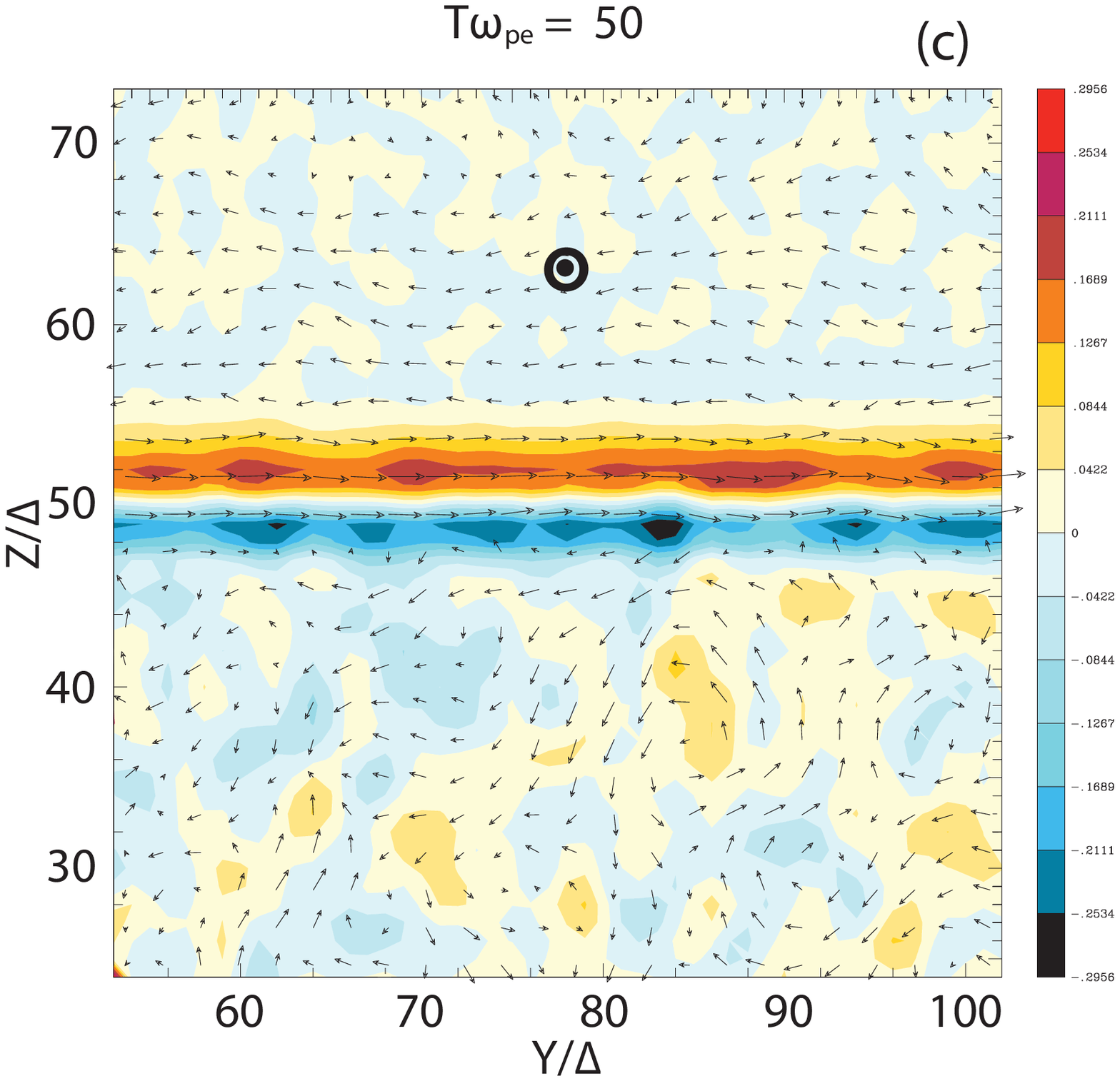} }
\,
\resizebox{0.85\hsize}{!}{\includegraphics[width=18cm]{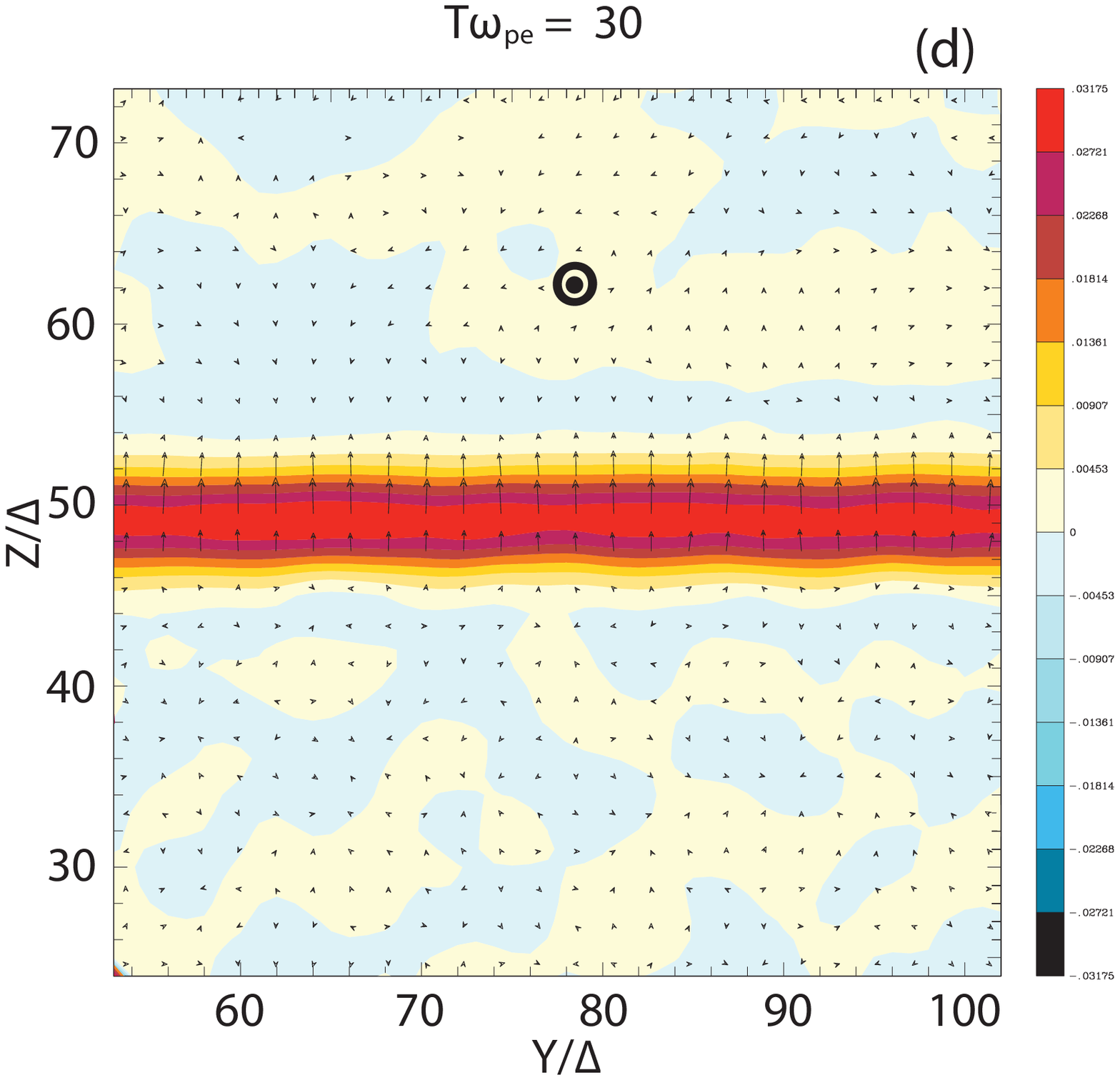}
\,
\includegraphics[width=18cm]{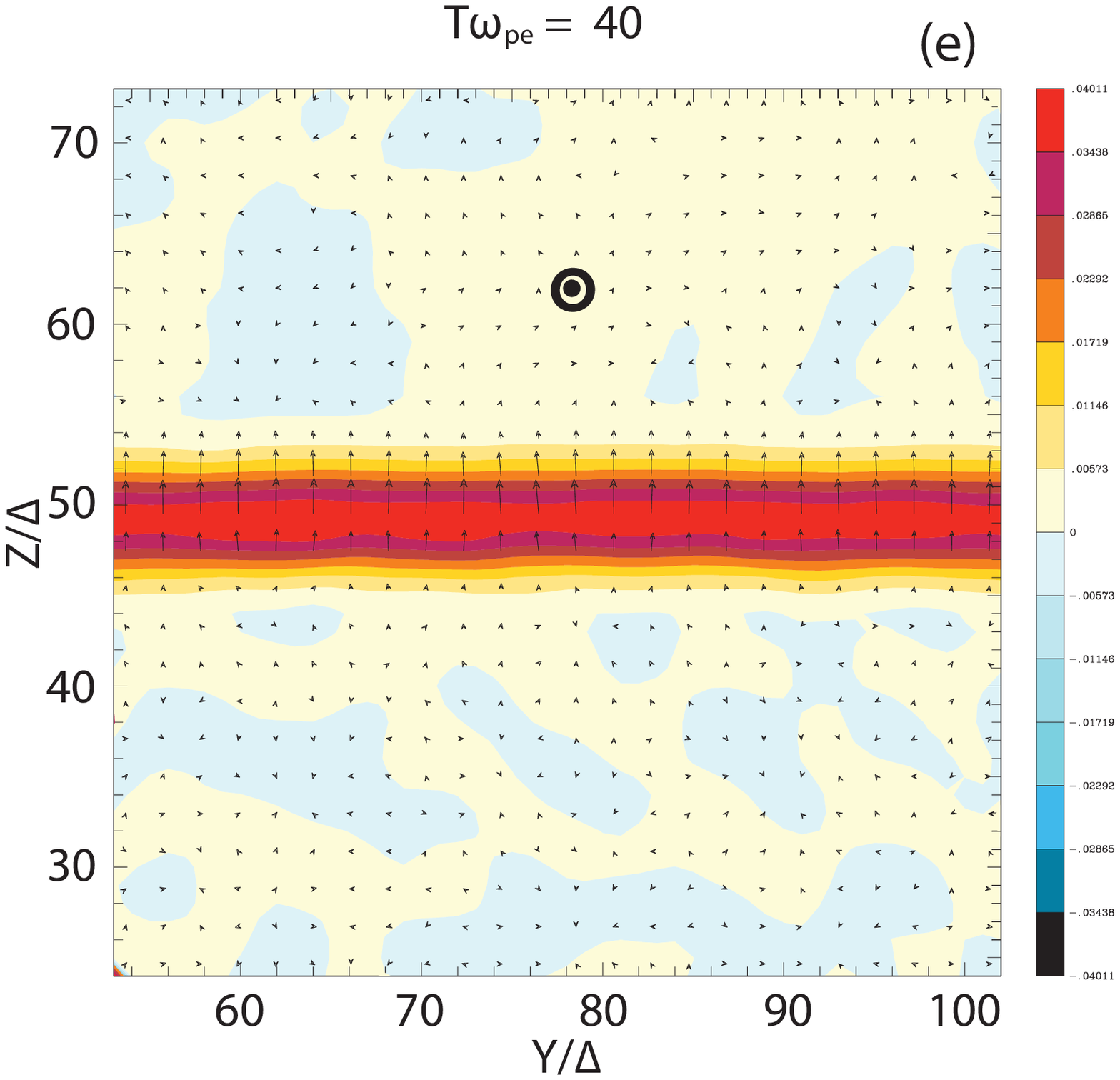}
\,
\includegraphics[width=18cm]{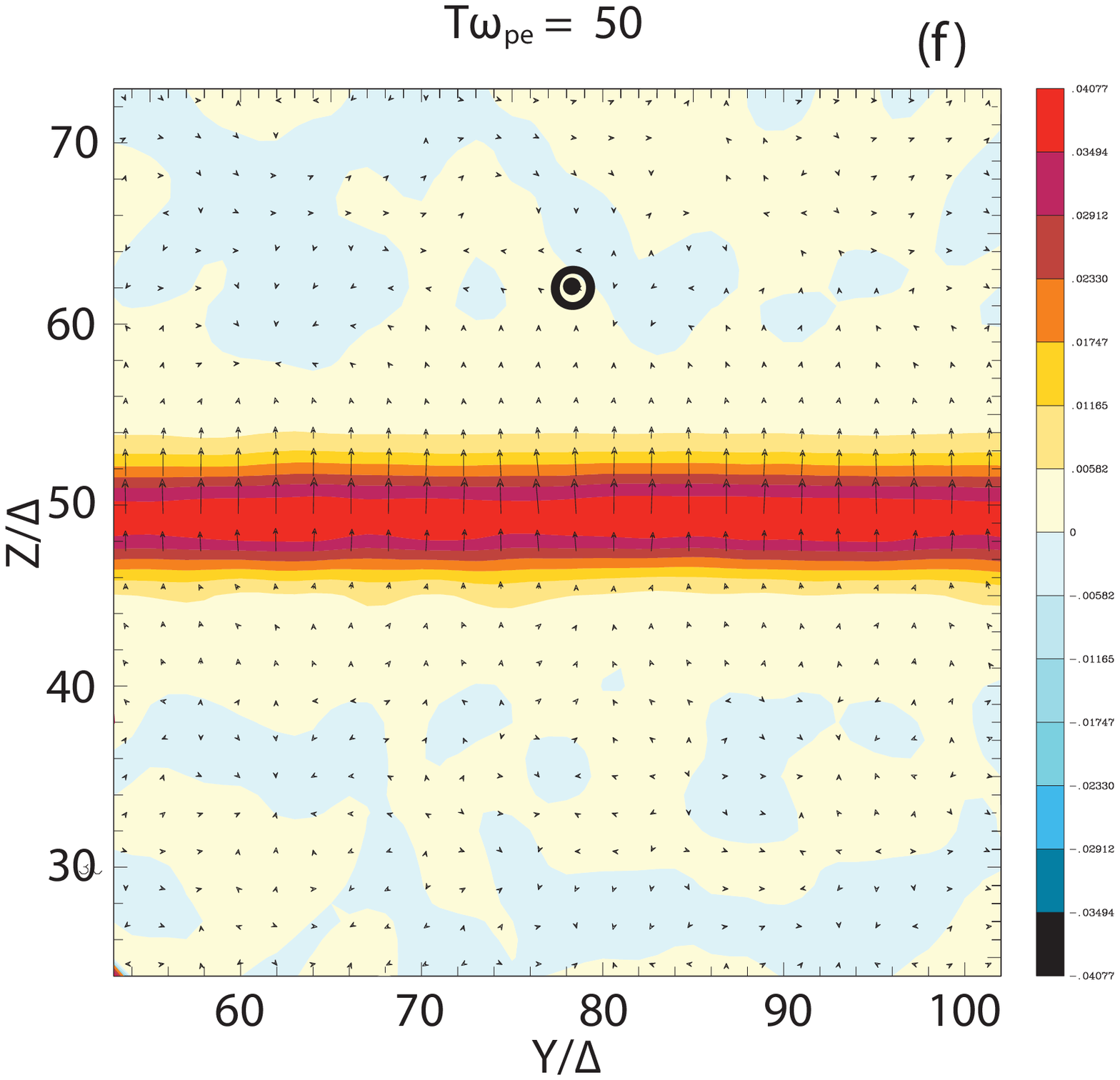} }
\end{center}
\vspace*{-0.5cm}
\caption{\baselineskip 12.0pt The time evolution of current filaments ($J_{\rm x}$) and electric field ($E_{\rm z}$) 
in the area  denoted by the small box in Fig. \ref{fig2}a.   KKHI starts to grow at (a) $t = 30 \omega_{\rm pe}^{-1}$ 
and current filaments become large but have not yet merged by (c) $t = 50 \omega_{\rm pe}^{-1}$. 
The maximum current density (simulation units) is (a) $\pm0.108$ at $t = 30 \omega_{\rm pe}^{-1}$, (b) $\pm 0.166$ 
at $t = 40 \omega_{\rm pe}^{-1}$, and (c) $\pm 0.296$ at $t = 50 \omega_{\rm pe}^{-1}$. The arrows show 
the magnetic field ($B_{\rm y}, B_{\rm z}$) (the length of the arrows are not scaled to the strength of the magnetic 
fields).  Panels (d), (e), and (f) show the time evolution of the electric field ($E_{\rm z}$)  in the same area as the current 
filaments. The rather uniform electric field grows even at $t = 30 \omega_{\rm pe}^{-1}$. The maximum
values of $E_{\rm z}$ change with time (d): $\pm$0.032, (e): $\pm$0.040, and (f): $\pm$0.040. As seen from green 
lines in Fig. \ref{fig4}a,  electric fields 
achieve a  maximum at  $t = 45 \omega_{\rm pe}^{-1}$ and subsequently decreases.
\label{fig3}}
\end{figure*}

Alves et al. (2012) found that  KKHI modulations are less noticeable in the relativistic regime because they are masked 
by a  strong DC magnetic field component  (negligible in the subrelativistic regime) with a magnitude greater 
than the AC component.  As the amplitude of the KKHI modulations grows the electrons from one flow cross 
the shear-surfaces and enter the counter-streaming flow. In their simulations the protons being heavier 
($m_{\rm p}/m_{\rm e} = 1836$)  are unperturbed. DC current sheets which point in the direction of the proton 
velocity form around the shear-surfaces. These DC current sheets induce a DC component in the magnetic field. 
The DC magnetic field is dominant in the relativistic scenario because a higher DC current is set up by the crossing 
of electrons with a larger initial flow velocity and also because the growth rate of the AC dynamics is lower by 
$\gamma_{\rm 0}^{3/2}$ compared with a subrelativistic case (Alves et al. 2013).  
It is very important to note  that this DC magnetic 
field is not captured in MHD (e.g., Zhang et al. 2009) or fluid theories because it results from intrinsically
kinetic phenomena. Furthermore, since the DC field is stronger than the AC field, a kinetic treatment is clearly 
required in order to fully capture the field structure generated in unmagnetized or weakly magnetized
relativistic flows with velocity shear. This characteristic field structure will also lead to a distinct radiation signature 
(Sironi \& Spitkovsky 2009; Martins et al. 2009; Frederiksen et al. 2010; Nishikawa et al. 2009b, 2010, 2011, 2012).

We previously reported results from our first simulations for a  core-sheath case with  $\gamma_{\rm j}=15$ and  $m_{\rm i}/m_{\rm e} = 20$ (Nishikawa et al. 2013). Here we report new simulation results using  the real mass ratio $m_{\rm i}/m_{\rm e} = 1836$. 
We  find some differences from previous counter-streaming cases.

\begin{figure*}[ht!]
\begin{center}
\resizebox{.55\hsize}{!}{\includegraphics[width=12cm]{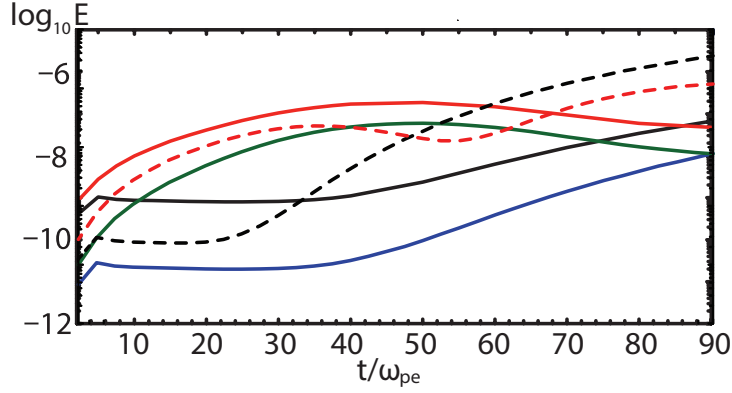}}
\end{center}
\vspace*{-0.7cm}
\caption{\baselineskip 12pt The panel show time evolution of electric field energy (red) and magnetic field
energy (black) for $\gamma_{\rm j}=15$ and  $m_{\rm i}/m_{\rm e} = 20$ (Nishikawa et al. 2013). For the case 
with the real mass ratio the green and blue lines shows the time evolution of electric field energy and magnetic 
field energy, respectively. The dashed red and black lines show the time evolution of electric field energy and 
magnetic field energy, respectively for the case with  $\gamma_{\rm j}=1.5$ and  $m_{\rm i}/m_{\rm e} = 20$. 
For all cases the kinetic KHI grows quickly,  the magnetic field energy plateaus, and later gradually grows again. 
In the non-relativistic case the magnetic field  energy becomes larger than the electric field
energy at an earlier time. \label{fig4}}
\end{figure*}

\begin{figure*}[ht]
\begin{center}
\resizebox{1.0\hsize}{!}{\includegraphics[width=20cm]{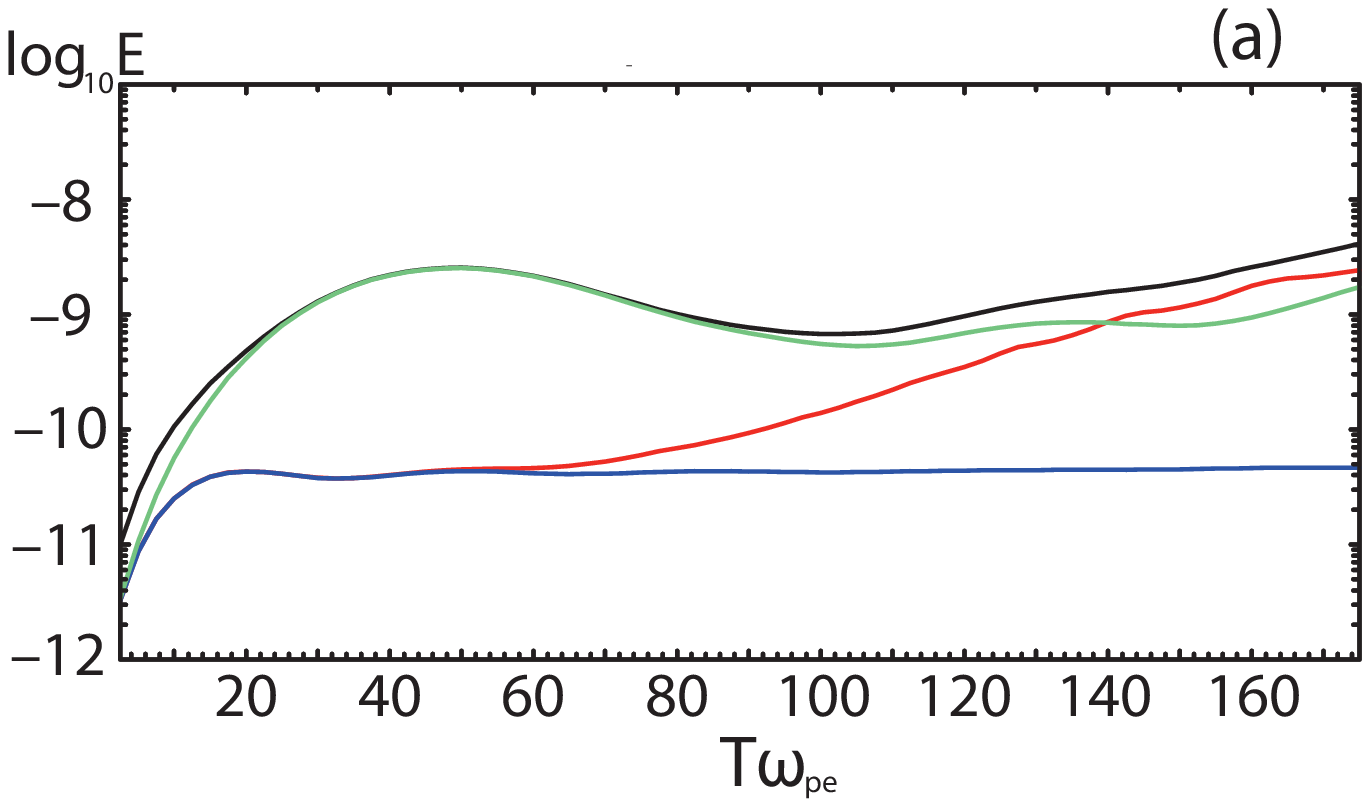}
\,
\includegraphics[width=20cm]{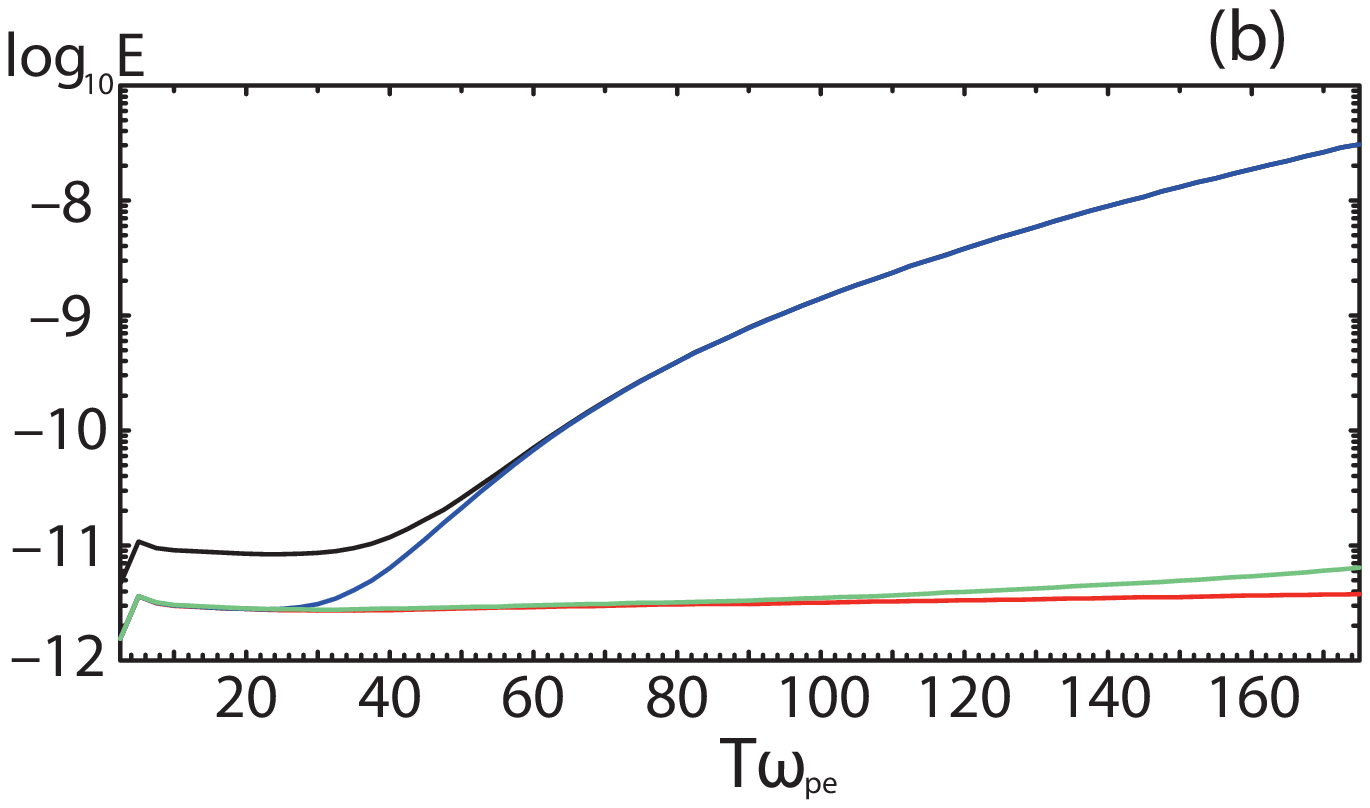}}
\end{center}
\vspace*{-0.3cm}
\caption{\baselineskip 12.0pt Evolution of  electric (a) and magnetic (b) field  energies  for the  core jet with 
with $\gamma_{\rm j} = 15$ and stationary sheath plasma for the case with the mass ration $m_{\rm i}/m{\rm e} = 1836$. 
Panel (a) shows the evolution of the total electric field energy (black), $E_{\rm z}$ (green), $E_{\rm x}$ (red), and
$E_{\rm y}$ (blue). The evolution of the total magnetic  field energy (black), $B_{\rm y}$ (blue), $B_{\rm z}$ (green),
and $B_{\rm x}$ (red) is plotted in panel (b). 
\label{fig5}}
\end{figure*}

\section{Magnetic and Electric Field Generation by  KKHI}
\subsection{Initial Core-Sheath Jet Conditions}

In this simulation study we investigate velocity shear in a core-sheath jet instead of the counter-streaming jets  used in 
 previous simulations (Alves et al. 2012; Liang et al. 2013).  
In this RPIC simulation the velocity shear occurs at the edges of  a velocity field with $v_{\rm core}$  pointing in the positive  $x$  direction  in the middle of the simulation box, with upper and lower quarters of the simulation box containing a velocity field with $v_{\rm sheath}$ pointing in the positive $x$ direction as indicated by the arrows in Figure \ref{fig1}a. Initially, the system is charge and current neutral. 

We have performed a simulation using a system with $(L_{\rm x}, L_{\rm y}, L_{\rm z}) = (1005\Delta, 205\Delta, 205\Delta)$ and with an ion to electron mass ratio of  $m_{\rm i}/m_{\rm e} = 1836$ ($\Delta$ is the cell size). 
The jet and sheath plasma density is $n_{\rm jt}= n_{\rm am} = 8$.  The electron skin depth, $\lambda_{\rm s} =
c/\omega_{\rm pe} = 12.2\Delta$, where $\omega_{\rm pe} = (e^{2}n_{\rm a}/\epsilon_0 m_{\rm e})^{1/2}$ is the electron
plasma frequency and the electron Debye length $\lambda_{\rm D}$ is  $1.2\Delta$.    The jet Lorentz factor  is
$\gamma_{\rm j} = 15$.  The jet-electron thermal velocity is
$v_{\rm j,th,e} = 0.014~c$ in the jet reference frame, where $c$ is the speed of light.  The electron/ion thermal velocity in
the ambient plasma is $v_{\rm a,th,e} = 0.03~c$.  Ion thermal velocities are smaller by $(m_{\rm i}/m_{\rm e})^{1/2}$.
We use periodic boundary conditions on all
 boundaries (Buneman 1993; Nishkawa et al. 2009).

\vspace{-0.cm}
\subsection{Core-Sheath Jet KKHI Results}
\vspace{-0.cm}

Our new simulation results using the real mass ratio are shown to compare with our previous results (Nishikawa et al. 2013).  

Figure \ref{fig1}b shows the magnitude of $B_{\rm y}$   plotted in the $x - z$ plane (jet flow in the $+x$-direction indicated by the large arrow) at the center of the simulation box, $y = 100\Delta$ at simulation time $t = 70\,\omega_{\rm pe}^{-1}$ for the case of  $\gamma_{\rm j}=15$ and $m_{\rm i}/m_{\rm e} = 1836$.

Figure \ref{fig2}a shows the magnitude of $E_{\rm z}$  plotted in the $y - z$ plane (jet flow is out of the page) at the center of the simulation box, $x = 500\Delta$ at $t = 30\,\omega_{\rm pe}^{-1}$.  Figure \ref{fig2}b   shows $E_{\rm z}$ (blue), $E_{\rm x}$ (black), and $E_{\rm y}$ (red) electric field components at $x = 500\Delta$ and $  y = 100\Delta$ at time  $t = 30\,\omega_{\rm pe}^{-1}$. 
As shown in Fig. \ref{fig4}, the $z$ component of electric field, $E_{\rm z}$ grows quickly, this DC electric field is shown
in Figs. \ref{fig2}a and \ref{fig2}b. As expected, the $y$ component of magnetic field , $B_{\rm y}$ is very small at this 
earlier time. The magnetic field grows later.   Figure \ref{fig2}c   shows $B_{\rm y}$ (red), $B_{\rm x}$ (black), and 
$B_{\rm z}$ (blue) magnetic field components at $x = 500\Delta$ and $y = 100\Delta$ $t = 70\,\omega_{\rm pe}^{-1}$. 

Figure \ref{fig3} shows the evolution of current filaments along the jet ($J_{\rm x}$) and the $z$ component of the electric field ($E_{\rm z}$) at an earlier time.  Jet flow  out of the page is  indicated by $\odot$.  Current filaments inside the jet  ($z/\Delta  > 50$) have positive $J_{\rm x}$ and outside the current filaments have  negative $J_{\rm x}$.  The width of both positive and negative current filaments is about 2 - 3 electron Debye lengths. This suggest that the jet electrons near the
velocity shear move out of the jet slightly and/or the sheath electrons are dragged by the jet. The steady current filaments at the
velocity shear create a steady $E_{\rm z}$ component. This electric field component decreases after  $t = 50\,\omega_{\rm pe}^{-1}$. These current filaments create a  $B_{\rm y}$ component which has been reported on in the previous work (Nishikawa et al.
2013).

We have compared the differences between  cases with  mass ratios $m_{\rm i}/m_{\rm e} = 20\; {\rm and}\; 1836$ 
for the relativistic jet  with $\gamma_{\rm j}=15$. We  find that the structure and growth rate of kinetic KHI is very similar 
(see Fig. \ref{fig4}). The heavier ions in the real mass ratio case  keep the system thermal fluctuations smaller, but the kinetic KHI grows similarly. The magnetic  field energy becomes larger than the electric field energy at a similar time in both cases around 
 $t = 87\,\omega_{\rm pe}^{-1}$. We  also performed a simulation with  $\gamma_{\rm j}=1.5$ and 
 $m_{\rm i}/m_{\rm e} = 20$.  For this non-relativistic case the magnetic field grows earlier and  overtakes the electric 
 field energy at  $t = 46\,\omega_{\rm pe}^{-1}$.

Figure \ref{fig5} shows the evolution of  electric and magnetic  field energy for     core jet  
with $\gamma_{\rm j} = 15$ and stationary sheath plasma for mass ratio $m_{\rm i}/m{\rm e} = 1836$.
Figure \ref{fig5}a shows  the evolution of the total electric field energy (black), $E_{\rm y}$ (green), $E_{\rm x}$ (red), and
$E_{\rm x}$ (blue). The $z$ component grows first as shown in Fig. \ref{fig3} and the $y$ component of the magnetic field
grows as shown in Fig. \ref{fig5}b. After $B_{\rm y}$ grows an  induced electric field $E_{\rm x}$ grows as
indicated by the red line in Fig. \ref{fig5}a. It should be noted that the  magnetic field grows at a slower rate
than  a counter-streaming case with  $\gamma_{\rm 0}=3.0$.  At  $t = 170\,\omega_{\rm pe}^{-1}$, saturation has not yet occurred.
This may be due 
our highly relativistic jet core resulting in a larger Lorentz factor in the ``electrostatic'' relativistic jet plasma frequency and larger electron inertia in the simulation frame.


\section{Core-Sheath Dispersion Relation}

We have extended the analysis presented in Gruzinov (2008) to core-sheath electron-proton plasma flows allowing for different core and sheath electron densities $n_{\rm jt}$ and $n_{\rm am}$, respectively, and core and sheath electron velocities $v_{\rm jt}$ and $v_{\rm am}$, respectively. In this analysis the protons are considered to be free-streaming whereas the electron fluid quantities and fields are linearly perturbed. We consider electrostatic modes along the jet.
The dispersion relation becomes:
\begin{eqnarray}
& &(k^{2}c^{2} + \gamma^{2}_{\rm am}\omega^{2}_{\rm p,am} - \omega^{2})^{1/2} (\omega - kV_{\rm am})^{2} \nonumber \\
&  &\times  [(\omega - kV_{\rm jt})^{2} - \omega^{2}_{\rm p,jt}] \nonumber \\
&  &+ (k^{2}c^{2} + \gamma^{2}_{\rm jt}\omega^{2}_{\rm p,jt} - \omega^{2})^{1/2} (\omega - kV_{\rm jt})^{2}\nonumber \\
& & \times   [(\omega - kV_{\rm am})^{2} - \omega^{2}_{\rm p,am}]  = 0 ,
\end{eqnarray}
where $\omega_{\rm p,jt}$ and  $\omega_{\rm p,am}$ are the plasma frequencies ($\omega^{2}_{\rm p} \equiv 4\pi n e^{2}/\gamma^{3}m_{\rm e}$) of jet and ambient electrons, respectively, $k = k_{\rm x}$
is the wave number parallel to the jet flow, and $\gamma_{\rm jt}$ and  $\gamma_{\rm am}$  are Lorentz factors of jet and ambient electrons, respectively.

Analytic solutions to equation (1) are not available except in the low ($\omega \ll \omega_{\rm p}$ and $kc \ll \omega_{\rm p}$) and high ($\omega \gg \omega_{\rm p}$ and $kc \gg \omega_{\rm p}$)  frequency and 
wavenumber limits.  In the high wavenumber limit with $V_{\rm am} =0$. In the high wavenumber limit $\omega \simeq kV_{\rm jt} \pm \omega_{\rm p,jt}/\sqrt 2$. In the low wavenumber limit 
an analytic solution to the dispersion relation is expressed as following:
\begin{eqnarray}
\omega & \simeq & {(\gamma_{\rm am}\omega_{\rm p,jt}kV_{\rm am} + \gamma_{\rm jt}\omega_{\rm p,am}kV_{\rm jt}) \over (\gamma_{\rm am}\omega_{\rm p,jt} +  \gamma_{\rm jt}\omega_{\rm p,am})} \nonumber \\
& \pm  & i {(\gamma_{\rm am}\omega_{\rm p,jt} \gamma_{\rm jt}\omega_{\rm p,am})^{1/2} \over (\gamma_{\rm am}\omega_{\rm p,jt} +  \gamma_{\rm jt}\omega_{\rm p,am})}k(V_{\rm jt}-V_{\rm am}) . \label{lowf}
\end{eqnarray}
Here the real part gives the phase velocity and the imaginary part gives the temporal growth
rate and directly shows the dependence of the growth rate on the velocity difference across the
shear surface. Equation (2) shows that the wave speed increases and the temporal growth rate decreases as
$\gamma_{\rm jt}\omega_{\rm p,am}/\omega_{\rm p,jt} = \gamma^{5/2}
_{\rm jt} (n_{\rm am}/n_{\rm jt})^{1/2}$ increases. 

We show exact solutions to Equation (2) in Figure (6) for a test case using the parameters, $n_{\rm jt} = n_{\rm am} = 0.1$ cm$^{-3}$, $V_{\rm jt}/c = 0.979796$ with $\gamma_{\rm jt} = 5.0$, and $V_{\rm am} = 0$.
\begin{figure}[h!]
\vspace{-0.3cm}
\begin{center}
\resizebox{0.97\hsize}{!}{\includegraphics[width=18cm]{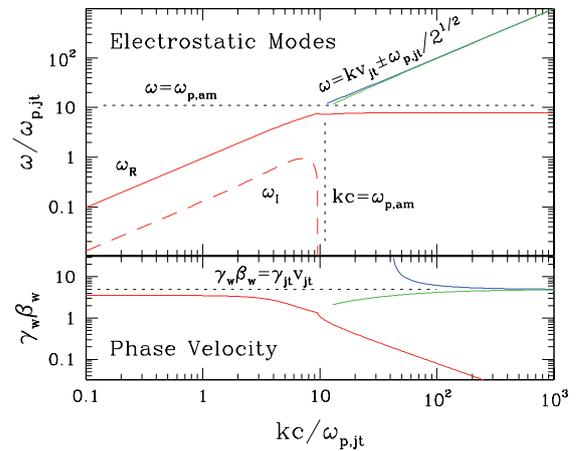}}
\end{center}
\vspace*{-0.7cm}
\caption{\baselineskip 12pt  The upper panel shows electrostatic mode solutions as a function of wavenumber, k, parallel to the flow direction. The real part, $\omega_{\rm R}$, and imaginary part, $\omega_{\rm I}$, of the frequency is indicated by the solid and dashed lines, respectively. The lower panel shows the phase velocity, $\gamma_{\rm w} \beta_{\rm w}$, where $ \beta_{\rm w} = \omega_{\rm R}/kc$. } 
\label{DRS}
\end{figure}
Note that the number densities are determined in their proper frames.  In the ambient observer frame the jet's number density would appear to be $n'_{\rm jt} = \gamma_{\rm jt} n_{\rm jt} = 0.5$ cm$^{-3}$. With the number densities determined in the proper frames the plasma frequencies are $\omega_{\rm p,jt} = (4 \pi n_{\rm jt} e^2/ \gamma_{\rm jt}^3 m_{\rm e})^{1/2} = 1.60 \times 10^3$ rad s$^{-1}$ and $\omega_{\rm p,am} = (4 \pi n_{\rm am} e^2/ \gamma_{\rm am}^3 m_{\rm e})^{1/2} = 1.78 \times 10^4$ rad s$^{-1}$. At small and large wavenumbers the numerical solutions follow the analytic low and high wavenumber solutions almost exactly.  The maximum temporal growth rate and associated wavenumber found from the numerical solutions shown in Figure 6 are $\omega^{\max}_{\rm I} \simeq 0.93 \omega_{\rm p,jt}$ at $k^{\max}c \simeq 7.0 \omega_{\rm p,jt}$ or $k^{\max}V_{\rm jt} \simeq 6.9 \omega_{\rm p,jt}$. 

If we assume that $\omega^{\max}_{\rm I} \sim \omega_{\rm p,jt}$ then we can use the imaginary part of eq.(2) to estimate the wavenumber at which temporal growth is a maximum with result that we estimate $k^{\max}V_{\rm jt} \sim 7.6 \omega_{\rm p,jt}$ and these estimates are within 10\% of the exact numerical solution.   The low wavenumber solution, eq.(2), provides excellent estimates of the complex frequency to wavenumbers within a factor of 2 of the maximally growing  wavenumber.



%

\section{Summary and Discussion}  

We have investigated generation of magnetic fields associated with velocity shear  between an unmagnetized relativistic jet and  an unmagnetized sheath plasma (core jet-sheath configuration).  We have examined the evolution of  electric and  magnetic fields generated by kinetic shear (Kelvin-Helmholtz) instabilities. Compared to the previous studies using counter-streaming performed by Alves et al. (2012), the structure of KKHI  of our jet-sheath configuration is slightly different even for the global evolution of the strong transverse magnetic field. We find that the major components of growing  modes are $E_{\rm z}$ and $B_{\rm y}$. After the $B_{\rm y}$ component is excited, the induced electric field $E_{\rm x}$ becomes larger. However, other components are very small. We find that the structure and growth rate with electron KKHI with the cases to the real mass ration $m_{\rm i}/m_{\rm e} = 1836$ and $m_{\rm i}/m_{\rm e} = 20$ are similar. In our simulations with jet-sheath case no saturation at the later time is   seen as in the counter-streaming cases.  This difference seems come from that fact that the jet is highly relativistic
and  our simulation is done in jet-sheath configuration. The growth rate with mildly-relativistic jet case ($\gamma_{\rm j} = 1.5$) is larger than the relativistic jet case ($\gamma_{\rm j} = 15$).

We have calculated, self-consistently, the radiation from electrons accelerated in
the turbulent magnetic fields in the relativistic shocks. We found that the synthetic spectra depend on the
Lorentz factor of the jet, the jet's thermal temperature, and the strength of the generated
magnetic fields (Nishikawa et al. 2011, 2012).  In forthcoming work we will obtain synthetic spectra from particles accelerated by KKHI as we have done for shock simulations \citep{nishikawa11,nishikawa12}.

\vspace{0.5cm}

This work is supported by NSF AST-0908010, and AST-0908040, NASA-NNG05GK73G,
NNX07AJ88G, NNX08AG83G, NNX08 AL39G,  NNX09AD16G, and NNX12AH06G. JN is supported by NCN through grant DEC-2011/01/B/ST9/03183. YM is supported by the Taiwan National Science Council under the grant
NSC 100-2112-M-007-022-MY3. Simulations were performed at the Columbia and Pleiades facilities at the
NASA Advanced Supercomputing (NAS), Kraken and Nautilus at The National Institute for Computational Sciences (NICS),
and Stampede at The Texas Advanced Computing Center (TACC) which are supported by the NSF. This research was started during the program ``Chirps, Mergers and Explosions:  The Final Moments of Coalescing Compact Binaries'' at the Kavli Institute for Theoretical Physics which is supported by  the National Science Foundation under Grant No. PHY05-51164.

\end{document}